\documentclass[%
 reprint,
superscriptaddress,
nofootinbib,
 amsmath,amssymb,
prd,
]{revtex4-1}

\usepackage{graphicx}% Include figure files
\usepackage{dcolumn}% Align table columns on decimal point
\usepackage{bm}% bold math
\usepackage[colorlinks]{hyperref}% add hypertext capabilities
\usepackage{color}
\usepackage{xspace}
\usepackage[caption=false]{subfig}
\def\beq{\begin{eqnarray}}
\def\eeq{\end{eqnarray}}

\usepackage[normalem]{ulem}

\newcommand{\Mpch}{h^{-1}\mathrm{Mpc}}
\newcommand{\hMpc}{h\,\mathrm{Mpc}^{-1}}
\newcommand{\hun}{\,\mathrm{km}\,\mathrm{s}^{-1}\mathrm{Mpc}^{-1}}
\newcommand{\lcdm}{$\Lambda$CDM\xspace}
\newcommand{\lya}{Ly$\alpha$\xspace}

\definecolor{darkgreen}{RGB}{0,120,0}
\definecolor{brown}{RGB}{120,60,0}

\newcommand{\new}[1]{#1}%\textcolor{brown}{#1}}
\newcommand{\resub}[1]{{#1}}
\newcommand{\neww}[1]{#1}%\textcolor{brown}{#1}}
\newcommand{\newww}[1]{#1}%\textcolor{brown}{#1}}

\begin{document}

%\preprint{APS/123-QED}

\title{Determining the Hubble Constant without the Sound Horizon: Measurements from Galaxy Surveys}% Force line breaks with \\

\author{Oliver H.\,E. Philcox}
\email{ohep2@cantab.ac.uk}
\affiliation{Department of Astrophysical Sciences, Princeton University,\\ Princeton, NJ 08540, USA}
\affiliation{Department of Applied Mathematics and Theoretical Physics, University of Cambridge,\\ Cambridge CB3 0WA, UK}%
\affiliation{School of Natural Sciences, Institute for Advanced Study, 1 Einstein Drive,\\ Princeton, NJ 08540, USA}
%\affiliation{Max-Planck-Institut für Astrophysik,\\ Karl-Schwarzschild-Str.~1, 85741 Garching, Germany}%
\author{Blake D. Sherwin}%
%\email{bds30@damtp.cam.ac.uk}
\affiliation{Department of Applied Mathematics and Theoretical Physics, University of Cambridge,\\ Cambridge CB3 0WA, UK}
\affiliation{Kavli Institute for Cosmology, Institute of Astronomy, University of Cambridge,\\ Cambridge CB3 0HA, UK}%
\author{Gerrit S. Farren}%
\affiliation{Department of Applied Mathematics and Theoretical Physics, University of Cambridge,\\ Cambridge CB3 0WA, UK}
\affiliation{Department of Physics and Astronomy, Haverford College,\\ Haverford, PA 19041, USA}
\author{Eric J. Baxter}%
\affiliation{Institute for Astronomy, University of Hawai’i,\\ 2680 Woodlawn Drive, Honolulu, HI 96822, USA}
%\affiliation{\eric{to write}}

%\author{\textit{et al.}}


\begin{abstract}
Two \resub{sources of geometric information} are encoded in the galaxy power spectrum: the sound horizon at recombination and the horizon at matter-radiation equality. \resub{Analyzing the} BOSS DR12 galaxy power spectra using \resub{perturbation theory} with $\Omega_m$ priors from Pantheon supernovae \resub{but no priors on $\Omega_b$}, we obtain constraints on $H_0$ from \resub{the} second scale, finding $H_0 = 65.1^{+3.0}_{-5.4}\,\mathrm{km}\,\mathrm{s}^{-1}\mathrm{Mpc}^{-1}$; this differs from the best-fit of SH0ES at 95\% confidence. Similar results are obtained if $\Omega_m$ is constrained from uncalibrated BAO: $H_0 = 65.6^{+3.4}_{-5.5}\,\mathrm{km}\,\mathrm{s}^{-1}\mathrm{Mpc}^{-1}$. Adding the analogous lensing results from Baxter \& Sherwin 2020, the posterior shifts to $70.6^{+3.7}_{-5.0}\,\mathrm{km}\,\mathrm{s}^{-1}\mathrm{Mpc}^{-1}$. Using mock data, Fisher analyses, and scale-cuts, we demonstrate that our constraints do not receive significant information from the sound horizon scale. Since many models resolve the $H_0$ controversy by adding new physics to alter the sound horizon, our measurements are a consistency test for standard cosmology before recombination. A simple forecast indicates that such constraints 
could reach $\sigma_{H_0} \simeq 1.6\,\mathrm{km}\,\mathrm{s}^{-1}\mathrm{Mpc}^{-1}$ in the era of Euclid.
\end{abstract}

\maketitle

%\tableofcontents

\section{Introduction}\label{sec: intro}
How do galaxy surveys measure the Hubble constant? \new{Recent} analyses have determined $H_0$ by comparing the angular scale of Baryon Acoustic Oscillations (BAO) with the theoretical size of the sound horizon scale at decoupling, $r_d$. A second `standard ruler' exists however; the equality scale, \textit{i.e.}, the horizon wavenumber %\eric{is light horizon correct here?}
at matter radiation equality, \resub{whose angular scale can be measured from the power spectrum shape and physical scale predicted by theory}. %, $k_\mathrm{eq}^{-1}$. %\blake{Eric and I had discussions about whether we can call keq a standard ruler in our previous paper and decided against it, because much of the information comes via the ruler itself, rather than the measured angular scale. On the other hand, it is quite a succinct way of explaining things, which I like. Keep or not? Let's discuss.}\oliver{I think it's useful for general interpretability}. 
In this work, we explore the extent to which galaxy surveys can use this scale to place constraints on $H_0$ that are independent of \newww{the sound horizon}.

Until recently, precise $H_0$ constraints have stemmed from two sources:
%Obtaining precise measurements of $H_0$ has been a central area of research over the past decade, with constraints historically stemming from two sources: 
the Cepheid-calibrated local distance ladder \citep[e.g.,][]{2019ApJ...876...85R,2020AJ....160...71S} and anisotropies in the cosmic microwave background (CMB) \citep[e.g.,][]{2020A&A...641A...6P}. Today, a host of additional constraints are available, arising from data-sets such as galaxy and Lyman-alpha (\lya) BAO \citep[e.g.,][]{2017MNRAS.470.2617A,2019JCAP...10..044C,2018MNRAS.480.3879A,2019ApJ...874....4A,2019JCAP...10..029S,2018ApJ...853..119A,2020arXiv200708991E}, strong gravitational lensing \citep[e.g.,][]{2019MNRAS.498.1420W,2020arXiv200702941B}, and gravitational wave observations
\citep{2017Natur.551...85A,2019NatRP...2...10R}. 
Broadly, these fall into two camps: `indirect' measurements, which require a full cosmological model for interpretation; and `direct' probes, independent of early-universe physics.\footnote{These are sometimes classified as `early' and `late' measurements respectively, but the terminology can be \resub{confusing}, since an `early' measurement might not involve high-redshift data-sets.} Probes in the former category, including CMB and calibrated BAO, usually derive information from the sound-horizon at recombination,\footnote{For the purposes of this work, there is little difference between `recombination' and `decoupling'; we thus use the terms interchangeably.} calculated assuming \lcdm. Previously, a tension between direct and indirect measurements seemed apparent; however, \resub{this distinction has become less clear} with the latest results from the TRGB-calibrated distance ladder \citep{2019ApJ...882...34F}, strong-lensing \citep{2020arXiv200702941B} and \resub{recalibrated mega-maser results \citep{2020arXiv201001119B}}. Nevertheless, there remains significant disagreement between indirect probes and the SH0ES distance ladder measurements \citep{2019ApJ...876...85R}, reaching a significance of $\sim 5\sigma$ \citep{2019NatAs...3..891V}.

Two primary possibilities exist to resolve this: (a) unresolved systematics \citep[e.g.,][]{2020arXiv200710716E}; or (b) incompleteness of the cosmological model. For the latter, a wide variety of new-physics models have been proposed; many of these %most promising models 
resolve the tension by providing mechanisms to reduce the sound horizon at recombination. %However, 
As yet, there is no generally accepted solution.

Ref.\,\citep{2020arXiv200704007B} proposed a new method to shed light on the discrepancy, using CMB lensing to measure $H_0$ without the sound horizon scale; constraints were instead derived from the 
%data from CMB lensing to constrain $H_0$ by way of the 
angular equality scale, $L_\mathrm{eq}$. Being simply the projected wavenumber of modes entering the horizon at $z_\mathrm{eq}\sim 3400$, this produces a definitive feature in the convergence spectrum, \newww{and can be used as a standard ruler}. 
%, the projected wavenumber of modes entering the horizon at $z_\mathrm{eq}\sim 3400$.
%\equiv k_{\rm eq} \chi_*$, where $\chi_*$ is the distance to last scattering.}
%\eric{maybe state explicitly $L_{\rm eq}\ \equiv k_{\rm eq} \chi_*$}. 
%\neww{This scale can be straightforwardly inferred from the lensing power spectrum and hence be used as a standard ruler}. 
%Given $\Omega_m$, $L_\mathrm{eq}$ can be predicted from theory \eric{don't you need $\Omega_m$ and $h$ info to get $L_{eq}$  (or $\omega_m$ and $h$)?  Even that's not really enough, since  you also need $T_{\rm CMB}$ or some other measure of radiation density to get $z_{\rm eq}$.  Could maybe just simplify to "$L_{\rm eq}$ can be used as a standard ruler"}, thus
%. Since the physical scale $k_\mathrm{eq}^{-1}$ \new{can be predicted from theory (given $\Omega_m$)},
%is known \eric{the physical scale only gets calibrated with $\Omega_m$ info, so this seems a bit misleading}
%it can be used as a standard ruler. % to constrain $H_0$
Importantly, the equality scale is sensitive to different redshifts than those of CMB and BAO analyses ($z\sim 1100$).  
%those of BAO analyses, \new{which are mainly sensitive to $z\sim 1100$.}
%\eric{we haven't explicitly stated before this point that sound horizon comes mainly from $z \sim 1100$, so this point about different redshifts might be a bit confusing}. 
This yields an important test: inconsistency of equality- and recombination-based $H_0$ constraints would give evidence for physics beyond \lcdm operating \newww{at $z\gtrsim 10^3$}. %in the interim redshifts \eric{it's not just interim, right?  keq and rs weight redshifts $z > 3000$ differently as well}. %decade of scale factor prior to recombination.
Combining \textit{Planck} lensing with cosmological priors on $\Omega_m$ and $A_s$, Ref.\,\citep{2020arXiv200704007B} obtained $H_0 = 73.8\pm 5.1\,\hun$; unfortunately, the projected improvements from future surveys were modest \new{owing to the intrinsically large cosmic variance.}% of the 2D convergence field.}
%\eric{owing to cosmic variance of the power spectrum of the 2D convergence field}. 

Since the number of modes available to a 3D galaxy survey is typically much greater than for CMB lensing, one might expect stronger constraints on $H_0$ from this avenue: indeed, this was the primary source of $H_0$ information from galaxy surveys two decades ago (\citep[e.g.,][]{1997PhRvL..79.3806T,2001MNRAS.327.1297P} \resub{and Ref.\,\citep{2013MNRAS.429.1902P} for a more recent attempt}). In this work, we perform such a measurement with modern surveys.%describe how we 

\resub{This paper has the following structure. In Sec.\,\ref{sec: physics} we discuss the physics behind our approach, before considering the data-sets and analysis pipeline in Sec.\,\ref{sec: data}. Results are presented in Sec.\,\ref{sec: results}, both for the BOSS data and idealized mock catalogs, before we present a discussion in Sec.\,\ref{sec: discussion}.}

\section{Equality and The Sound Horizon}\label{sec: physics}
A glance at the matter power spectrum reveals two features: the broadband peak at wavenumber $k_\mathrm{eq}\sim 10^{-2}\hMpc$ and the oscillatory behavior with period $\Delta k \sim 0.05\hMpc$. The behavior around $k_\mathrm{eq}$ is well known, arising from the transition between modes that enter the horizon in radiation-dominated and matter-dominated epochs. %, \new{leading to a}
%\sout{and thus a}\eric{leading to a} 
%change in the transfer function. 
In galaxy surveys, resolving the peak is difficult (though possible with experiments such as SPHEREx \citep{2014arXiv1412.4872D}), due to relativistic effects and integral constraints \citep[e.g.,][]{2019JCAP...08..036D}, %\new{integral constraints (since the mean survey density is inferred from the survey itself)},
%\eric{what do you mean by normalization effects?}, 
alongside cosmic variance \resub{and imaging systematics}. More generally, $k_\mathrm{eq}$ information is encoded in the shape of the power spectrum, and can be inferred from smaller \resub{(though still linear)} scales,
%\bds{information is encoded in the power spectrum shape, and $k_\mathrm{eq}$ can be inferred even away from the peak, especially given additional amplitude and bias information from redshift-space distortions or priors).}
%\oliver{removed this scaling since it's inexact and takes space}
as seen by the approximate scaling solution to the linearly-biased perturbation equations:
\beq\label{eq: meszaros}
    P_g(k) \approx \begin{cases}b_1^2A_s\left(\frac{k}{k_\mathrm{eq}}\right)^{n_s} & k < k_\mathrm{eq}\\ b_1^2A_s\left(1+\log\left(\frac{k}{k_\mathrm{eq}}\right)\right)^2 \left(\frac{k}{k_\mathrm{eq}}\right)^{n_s-4} & k > k_\mathrm{eq}.\end{cases}
\eeq
This is facilitated in part by the addition of amplitude and bias information from redshift-space distortions or priors.\footnote{\resub{If the transfer function was a pure power-law we would expect full $A_s-k_\mathrm{eq}$ degeneracy; the logarithmic shape for $k>k_\mathrm{eq}$ reduces this, though we note that measuring $k_\mathrm{eq}$ in this manner is inherently model-dependent.}}

%\blake{are these equations useful? not totally sure}, \oliver{could remove these}
%for linear galaxy bias $b_1$, primordial tilt $n_s$ and amplitude $A_s$. 
%Given amplitude and bias information (from redshift-space distortions or priors), the shape of the power spectrum %(along with the amplitude, given some knowledge of $A_s$) \oliver{we actually need b1^2 A_s, which can be gotten directly from RSD}
%allows $k_\mathrm{eq}$ to be determined. 

In \lcdm, the equality scale is simply related to cosmological parameters;
\beq
    k_\mathrm{eq} = \left(2\Omega_{cb}H_0^2z_\mathrm{eq}\right)^{1/2},\,\, z_\mathrm{eq} = 2.5\times 10^{4}\Omega_{cb} h^2\Theta_{2.7}^{-4}
\eeq
\citep{1998ApJ...496..605E,2019JCAP...11..034C}, where $\Theta_{2.7}\equiv T_\mathrm{CMB}/(2.7\,\mathrm{K})$ is the temperature of the CMB monopole, $\Omega_{cb}\equiv \Omega_{cdm} + \Omega_b$ (assuming neutrinos to be relativistic at $z_\mathrm{eq}$) \resub{is the CDM+baryon density fraction} and $h\equiv H_0\,/\,(100\hun)$. Measuring $k_\mathrm{eq}$ in $\hMpc$ units probes the combination $\Omega_{cb} h\equiv (\omega_{cdm}+\omega_b)/h$, or, marginalizing over $\omega_{b}$, $\omega_{cdm}/h$. \neww{Given $k_\mathrm{eq}$ and a probe of $\Omega_{cb}$ \resub{(or, more commonly, $\Omega_m$)} we can thus solve for the Hubble constant.}

Complicating this is the second scale: %method for deriving $H_0$ 
%are the effects of the other scale: 
the sound horizon at $z_d$, the redshift of photon-baryon decoupling. This is given by
\beq\label{eq: r-d-approx}
    r_d \equiv r_s(z_d) &=& \int_{z_d}^{\infty} \frac{c_s(z)}{H(z)}dz\\\nonumber
    &\approx& \frac{55.154h\exp\left[-72.3(\omega_\nu+0.0006)^2\right]}{\omega_{cb}^{0.25351}\omega_b^{0.12807}}\,\Mpch
\eeq
\citep{2015PhRvD..92l3516A}, where $H(z)$ and $c_s(z)$ are the Hubble parameter and sound-speed. $r_d$ sources two main features: the BAO wiggles with $\Delta k \approx 0.05\hMpc$, and a small-scale suppression of power \newww{on the baryonic Jeans scale} \citep{2006PhR...429..307L}. Both have amplitudes scaling as $\omega_b/\omega_{cb}$ and could be used to infer the \new{physical scale} of the sound horizon. % (assuming standard cosmology), 
\neww{In combination with the measured angular scale, this constrains $H_0$.} %\blake{Isn't the worry only that the depth of the small scale suppression calibrates rd? Would not say both...}
%\eric{maybe: the angular scale of the sound horizon}.
From \eqref{eq: r-d-approx}, such measurements carry the degeneracy $\omega_{cdm}\propto h^{4}$ (after $\omega_b$ marginalization), and a measurement using the Jeans suppression is degenerate with $n_s$.  %\eric{since this is the first time introducing baryonic suppression, it might be good to be more explicit about what you mean by "Jeans suppression"}
%\blake{An important, basic discussion I am missing here is an outline of how we get from a measurement of the equality scale to a constraint on Hubble. Just keq/h is Omegamh, with a probe of Omegam i can solve for h. This illustrates the importantce of an Omegam prior as well.} %\blake %Note that these effects are far less prevalent in lensing-based measurements, due to the integrated nature of the statistic.
%\eric{it might also be useful to explicitly point out in this section how the angular scale of $r_d$, combined with the calibrated physical scale of $r_d$ combine to constrain $h$.}

An $H_0$ measurement from the full-shape (FS) of the galaxy power spectrum will include information from both $r_d$ and $k_\mathrm{eq}^{-1}$ standard rulers, whilst BAO analyses are sensitive only to $r_d$. To extract only information deriving from equality, %-based \neww{$H_0$} information,
%\eric{about $H_0$,} 
one may wish to `marginalize over the sound horizon'; this is non-trivial 
%not really possible \oliver{it is possible *effectively* e.g. with the covariance thing}
since $r_d$ is not a direct input to any Boltzmann code, emerging only following simplifications such as tight-coupling.\footnote{An \textit{ad-hoc} rescaling of $r_d$ would be dangerous, as it would not conserve the stress-energy $T_{\mu\nu}$; instead, one should self-consistently add any new physics model to the perturbation equations.} Here, we limit the $H_0$ information arising from the sound horizon simply by removing the usual informative prior on $\omega_b$, and thus the external $r_d$ calibration \resub{even if $\omega_{cb}$ is known precisely}.
%can prevent $H_0$ information arising from the sound horizon by simply removing the usual informative prior on $\omega_b$, and thus the external $r_d$ calibration. 
For future data this may be insufficient: the BAO features and Jeans suppression can, in principle, calibrate each other, sourcing an effective sound-horizon prior.% on $r_d$.

\section{Data-sets and Analysis}\label{sec: data}
\subsection{Redshift-Space Power Spectrum}\label{subsec: pk-data}
%In this work, 
Our main observational data-set is the twelfth data release (DR12) \citep{2017MNRAS.470.2617A} of the Baryon Oscillation Spectroscopic Survey (BOSS), part of SDSS-III \citep{2011AJ....142...72E,2013AJ....145...10D}. Split across two redshift bins \resub{(at $z = 0.38, 0.61$)} in each of the Northern and Southern galactic caps, the survey contains $\sim 1.2\times 10^6$ galaxy positions with a total volume of $5.8h^{-3}\mathrm{Gpc}^3$. Here, we use the %publicly available 
(unreconstructed) power spectrum monopole and quadrupole,\footnote{\href{https://fbeutler.github.io/hub/boss_papers.html}{fbeutler.github.io/hub/boss\_papers.html}} each in 48 $k$-bins for $k\in[0.01,0.25]\hMpc$, with % between $0.01\hMpc$ and $0.25\hMpc$.
covariances generated from a suite of 2048 MultiDark-Patchy mocks \citep{2016MNRAS.456.4156K,2016MNRAS.460.1173R}, using 
%the which share the BOSS survey geometry and 
the cosmology $\{\Omega_m = 0.307115,\Omega_b = 0.048,\sigma_8 = 0.8288, h= 0.6777, \sum m_\nu = 0\,\mathrm{eV}\}$.\footnote{\resub{We caution that the spectra may have non-trivial imaging systematics at low-$k$; investigation of these is beyond the scope of this work.}} 

To extract maximal shape information, we model $P_\ell(k)$ with
%Since most $k_\mathrm{eq}$ information appears at large scales, it might seem natural to use linear theory to analyze the full shape of the power spectrum multipoles, $P_\ell(k)$. This requires harsh $k$-space cuts however, and may lead to information loss since one-loop features are known to tighten constraints on key parameters such as $\omega_{cdm}$. For full generality, we will use 
the Effective Field Theory of Large Scale Structure, following %model for $P_\ell(k)$ presented in
Ref.\,\citep{2020JCAP...05..042I} (see also Ref.\,\citep{2020JCAP...05..005D}). This includes %linear theory \blake{confusing to say linear theory - maybe just say 1 loop PT?},
one-loop perturbation theory, infra-red resummation of long-wavelength modes, and counterterms parametrizing the impact of small-scale physics. The model is convolved with the survey window function, and incorporates Alcock-Paczynski (AP) effects \citep{1979Natur.281..358A}.\footnote{We use a fiducial value of $\Omega_m = 0.31$ to apply the AP rescaling to the BOSS data.} % This effect is not present for mock catalogs.} 
The procedure has been used in a number of works \citep{2020PhRvD.101h3504I,2019JCAP...11..034C,2020JCAP...05..032P,2020arXiv200611235I,om4}, including a rigorous test on huge volume simulations \citep{2020arXiv200308277N}, \resub{showing any theory error to be strongly subdominant to the BOSS statistical error}. Here, we utilize the \texttt{CLASS-PT} implementation \citep{2020PhRvD.102f3533C}, with MCMC performed using \texttt{montepython} v3.3 alongside heavily optimized public likelihoods,\footnote{\href{https://github.com/michalychforever/lss_montepython}{github.com/michalychforever/lss\_montepython}.} \resub{with convergence assumed once the Gelman-Rubin diagnostic is below $1.05$}.
We vary the parameter set
\beq
    &&\{h, \omega_{cdm}, \omega_b, A_s/A_{s,\mathrm{Planck}}, n_s, \sum m_\nu\}\\\nonumber
    &&\times\quad\{b_1,b_2,b_{G_2},b_4,c_{s,0},c_{s,2},P_\mathrm{shot}\}
\eeq
where $A_{s,\mathrm{Planck}} = 2.0989\times 10^{-9}$, and $\sum m_\nu$ is the summed mass of three degenerate neutrinos. \newww{To aid convergence, we add a Gaussian prior on $\omega_b$ of width $50\%$ \resub{centered at the \textit{Planck} best-fit} and flat priors of $[0,0.18]\,\mathrm{eV}$ and $[0.87,1.07]$ to $\sum m_\nu$ and $n_s$ respectively.}\footnote{\resub{Initial testing showed that these do not significantly affect the $H_0$ constraints; indeed we can increase the $\omega_b$ prior width by a factor of $2$ without changing $\sigma_{H_0}$, though this leads to slower convergence. This is further discussed in Sec.\,\ref{subsec: mock-constraints}.}} The second line gives nuisance parameters of the EFT model, which are allowed to vary independently in each of the four data \neww{patches} \resub{necessitated by their differing redshifts and calibrations}, %chunks \blake{vary independently? also chunks sounds very informal and I also don't totally understand it, is there a better and clearer word?}, 
subject to the weak Gaussian priors of Ref.\,\citep{2020JCAP...05..042I}. Those entering the likelihood linearly ($b_4$, $c_{s,0}$, $c_{s,2}$ and $P_\mathrm{shot}$) are marginalized analytically \citep{2002MNRAS.335.1193B,2010MNRAS.408..865T,om4}, reducing the total number of %\eric{\sout{MCMC}} 
sampled parameters to $6 + 3\times 4 = 18$.

% some of which can be marginalized over analytically (and exactly)

% \oliver{HERE}
% In all analyses we vary the following parameters;
% \beq
%     &&\{h, \omega_{cdm}, \omega_b, A_s/A_{s,\mathrm{Planck}}, n_s, \sum m_\nu\}\\\nonumber
%     &&\times\quad\{b_1,b_2,b_{G_2},b_4,c_{s,0},c_{s,2},P_\mathrm{shot}\}
% \eeq
% where the fiducial amplitude $A_{s,\mathrm{Planck}} = 2.0989\times 10^{-9}$ is set to the \textit{Planck} value, and $\sum m_\nu$ is the sum of the neutrino masses.\footnote{We assume three degenerate massive neutrinos for this analysis, though this assumption is not expected to affect our analysis, given that the BOSS constraints on neutrino masses are weak \citep{2020PhRvD.101h3504I}.} The parameters in the second line are nuisance parameters including galaxy biases, ultraviolet counterterms and shot-noise. As noted in Refs.\,\citep{2002MNRAS.335.1193B,2010MNRAS.408..865T}, any parameters entering the model linearly can be marginalized over analytically and exactly. Here, we apply this to $b_{4}, c_{s,0},c_{s,2}$ and $P_\mathrm{shot}$, following Ref.\,\citep{om4}, assuming the same weak Gaussian priors. This reduces the number of free parameters $6+ 3\times 4 = 18$ (noting that the data patches differ in calibration and selection functions, thus must have separate free parameters). 

Later, we will require mock data \resub{to explore the information content of our model}.
%To test whether our model retains dependence on $r_d$, we consider mock data. 
This is generated from the theory model using the baseline cosmology $\{\omega_{cdm} = 0.118, \omega_b = 0.022, A_s/A_{s,\mathrm{Planck}} = 1.025, h = 0.6777, n_s = 0.9649, \sum m_\nu = 0.06\,\mathrm{eV}\}$, similar to the MultiDark-Patchy parameters, but with massive neutrinos. Three samples %ets \neww{of mock data} 
%\eric{of mock data} 
are created: (1) fiducial; (2) with negligible BAO wiggles; and (3) with fewer baryons (to reduce both BAO and baryon damping effects). For each, we fit nuisance parameters to the observed $P_\ell(k)$ and do not include \neww{noise}, \resub{such that all data-sets may be simply compared.}
%sample variance \blake{explain what you mean here? at least to me on the skype chat if I am being dumb!}, 
%thus providing a rigorous test of the analysis. 
Set (2) is generated by \neww{increasing the BAO damping scale by} %a factor of 
$1000\times$,
%in the theory model}, 
%using a very large \blake{numbers/details?} BAO damping scale in the theory model, 
whilst (3) reduces $\Omega_b$ by $10\times$ \neww{relative to the fiducial value}, %\blake{from the fiducial value?}, 
keeping %the physical parameters 
$\omega_{cdm}$ and $A_s$ fixed. \resub{When performing parameter inference, all mock data-sets are analyzed using the MultiDark-Patchy covariance matrix described above.}

% To test whether our model retains dependence on the sound horizon, it is useful to consider mock data. These are generated using the baseline cosmology $\{\omega_{cdm} = 0.118, \omega_b = 0.022, A_s/A_{s,\mathrm{Planck}} = 1.025, h= 0.6777, n_s = 0.9649, \sum m_\nu = 0.06\,\mathrm{eV}\}$; matching the MultiDark-Patchy $\Omega_m$, $\Omega_b$ and $\sigma_8$ but with the addition of massive neutrinos. We generate three data-sets: (1) a fiducial sample; (2) a sample without BAO wiggles; (3) a sample with the baryon density reduced by a factor of 10 (thus reducing both BAO wiggles and the baryon damping effect). For each, we generate spectra for each of the BOSS data chunks, and do not include noise, to elucidate any biases in the analysis. For each, bias parameters are generated by fitting to the observed power spectrum multipoles. The no-wiggle data-set can be simply generated by using a large BAO damping scale, $\Sigma_\mathrm{BAO}$, when generating the mock data, to suppress any wiggle contributions. For consistency, when analyzing these mocks, we use the same value of $\Sigma_\mathrm{BAO}$. To generate set (3), where $\omega_b$ is varied, we have two options: fix the physical parameters $\omega_{cdm}$ and $A_s$ or the derived parameters $\Omega_m$ and $\sigma_8$. We adopt the latter option, since the full-shape of the galaxy power spectra is known to be primarily sensitive to $\omega_{cdm}$ \citep{2019JCAP...11..034C}.

\subsection{Cosmological Priors}\label{subsec: cosmo-priors}
Equality based measurements of $H_0$ are assisted by information on $A_s$ (to constrain $k_\mathrm{eq}$) and $\Omega_{cb}$ (to break the $H_0-\Omega_{cb}$ degeneracy).\footnote{Knowledge of $b_1$ is also useful; this is provided by redshift-space distortions.} For the former, we employ a weak Gaussian prior of $A_s = (2.11\pm 0.36)\times 10^{-9}$, centered on the \textit{Planck} best-fit \citep{2020A&A...641A...6P}. The $r_d$-dependence of this is minimal, since the CMB measurement is limited by the optical depth and hence derives from very large scales; however, to be maximally conservative, we choose the prior width to be $10\times$ that of the \textit{Planck} constraint. %prior to have $10\times$ the width of the \textit{Planck} constraint in order to be maximally conservative. %but $10\times$ the width, to eliminate any residual $r_d$ information.

%To facilitate an equality-based $H_0$ constraint, it is useful to have additional information on $A_s$ (to improve the measurement of $k_\mathrm{eq}$ on scales away from the peak); \neww{most importantly, we require information on $\Omega_{cb}$ to break the degeneracy with $h$ arising from a measurement of the equality scale}.\footnote{We also need knowledge of $b_1$; this is provided by redshift-space distortions.} 
%\blake{a bit confusing to sometimes cay Omegacb and sometimes Omegam. Always use Omegam? or is there some subtlety with neutrino mass?} (to break the $\Omega_{cb}h$ degeneracy).

%For the former, a very weak $\sim 8\%$ prior centered on the Planck best-fit \citep{2018arXiv180706209P} is used; $A_s = (2.11\pm 0.18)\times 10^{-9}$. Whilst this is CMB-motivated, it is effectively independent of $r_d$ (as seen from the \textit{Planck} posterior contours), since it is limited by the optical depth $\tau$ \citep{2020arXiv200704007B}, measured from the largest scales, \neww{whilst the $r_d$ information arises from the smaller-scale peak locations.}
%\eric{while the information about $r_d$ comes from the peak locations}. 
%To be maximally conservative, we use a much broader prior than the $1.6\%$ posterior reported in Ref.\,\citep{2018arXiv180706209P}. To avoid unphysical regions of parameter space being explored, we additionally impose a $50\%$ prior on $\omega_b$, which helps with convergence and posterior Gaussianity, but does not add significant $r_d$ information (as we later demonstrate).%affect the $H_0$ constraints.

For the $\Omega_m$ prior, we principally use the marginalized result from Pantheon supernovae: $\Omega_m = 0.298\pm 0.022$ \citep{2018ApJ...859..101S}. 
%As in Ref.\,\citep{2020arXiv200704007B}, we primarily use the Pantheon supernova sample \citep{2018ApJ...859..101S} for the $\Omega_m$ prior, giving $\Omega_m = 0.298\pm 0.022$. 
This cannot constrain $H_0$ directly, since the supernova absolute magnitudes are unknown. An alternative source of $\Omega_m$ information is given by
%For some of our constraints, we also consider an alternative source of $\Omega_m$ information:
%method to probe $\Omega_m$: 
\textit{uncalibrated} BAO measurements. A standard BAO analysis proceeds by comparing the radial and angular oscillatory scales to the \lcdm sound horizon, %$r_d$, %through a measurement of the AP parameters. These
providing information on $\Omega_m$ and $H_0r_d$ through \newww{the evolution of} the angular diameter distance and Hubble parameter
%at the sample redshift, encoding $\Omega_m$ and $H_0r_d$
\citep[e.g.,][]{2019JCAP...10..044C}.
%, with strongest $\Omega_m$ constraints found at high redshift \citep{2019JCAP...10..044C}. 
To remove the dependence on $r_d$, %(and thus the $H_0$ constraining power, 
we rescale the sound horizon by a free parameter $\alpha_{r_d}$ \resub{when including BAO $H(z)r_d$ and $D_A(z)/r_d$ measurements}. In this formalism, no knowledge of recombination physics is required, just the existence of a time-independent correlation function peak. Here, we use a range of galaxy BAO measurements %stretching from $z = 0.106$ to $z = 2.34$, 
from BOSS DR7 \resub{(6dF\resub{GS} and Main Galaxy Samples)} \citep{2011MNRAS.416.3017B,2015MNRAS.449..835R}, and eBOSS DR14 Lyman-alpha measurements, (including cross-correlations with quasars) \citep{2019A&A...629A..85D,2019A&A...629A..86B}. We exclude the BOSS DR12 BAO measurements, since they are covariant with the FS data-set, which would cause additional complications. Alone, the uncalibrated BAO \resub{are found to} give the constraint $\Omega_m = 0.308^{+0.025}_{-0.030}$. \resub{For analyses using mock data, we center the priors on the true parameter values, keeping the same fractional width.}

\subsection{Additional Data-sets}\label{subsec: add-data}

$H_0$ constraints from CMB lensing were demonstrated in Ref.\,\citep{2020arXiv200704007B}. %These proceed by measuring of the angular equality scale $L_\mathrm{eq}$ %(analogous to the above discussion) 
Due to the presence of projection integrals, the measurements are relatively free from $r_d$-calibration, 
%baryonic effects \eric{by baryonic effects you mean $r_d$ calibration of $H_0$, right?  this might be confusing, as "baryonic effects" typically means impact of e.g. AGN on the power spectrum} 
\new{even with a restrictive prior on $\omega_b$.} % is introduced.}
%, especially without a restrictive prior on $\omega_b$ \eric{even with a restrictive prior on $\omega_b$, we found that $r_d$ information didn't inform Hubble}. 
The lensing power spectrum measures the combination $L_\mathrm{eq} \sim \Omega_m^{0.6}h$, with a different scaling than that of $k_\mathrm{eq}$, thus we may expect some degeneracy breaking \new{when this is combined with FS measurements.}
%\eric{you mean degeneracy breaking when combined with FS info right?}. 
Here, we use the public \textit{Planck} 2018 CMB-marginalized lensing likelihood \citep{2020A&A...641A...5P}, assuming zero covariance between this and the BOSS data.\footnote{Technically, some correlation will be present since the probes partially overlap. 
%and (c) the overlap is only partialBOSS only has partial overlap with the full \textit{Planck} footprint, 
Since the lensing kernel is much broader in redshift space than the BOSS selection function, and CMB lensing is only sensitive to %the small fraction of \oliver{this is not a small fraction. It's ~ 2/3}
modes that are perpendicular to the line-of-sight, we expect this to be small.} For analyses including the lensing data-set, we impose a twice tighter prior on $A_s$ of $(2.11\pm 0.18)\times 10^{-9}$ (as in Ref.\,\citep{2020arXiv200704007B}), to break the significant $A_s-L_\mathrm{eq}$ degeneracy.

%As shown in Ref.\,\citep{2020arXiv200704007B} we can additionally obtain recombination-independent bounds on cosmology from CMB lensing, via measurement of the equality scale $L_\mathrm{eq}$, in direct analogy to the above discussion for the galaxy power spectrum. Due to the projection integrals implicit in the statistic, the main baryon effects (and BAO wiggles) are washed out, thus we may be assured that little information is obtained from physics at the sound horizon, particularly given the absence of a strong prior on $\omega_b$. Unlike the BAO data, this can constrain $H_0$ directly, with the degeneracy direction $L_\mathrm{eq} \sim \Omega_m^{0.6}h = \mathrm{const.}$, somewhat different to that of the galaxy power spectrum. Here, we implement this using the public Planck 2018 lensing likelihoods, which are described in Ref.\,\citep{2019arXiv190712875P}. 

%In this analysis, we assume the CMB lensing measurements to be independent of the BOSS galaxy sample, \textit{i.e.} zero cross-covariance between the data-sets. In general, there will be some level of dependence, since the all-sky \textit{Planck} lensing results have an overlap of $\sim 23\%$ with the BOSS footprint. However, two effects act to reduce this correlation: (1) the CMB lensing kernel is very broad in redshift space, whereas BOSS measures only galaxies with $z \lesssim 0.7$; (2) CMB lensing is not sensitive to modes along the line of sight, unlike galaxy surveys. The combination of these will ensure that any cross-covariance is slight, and may be safely neglected.

\section{Results}\label{sec: results}

\begin{figure}
  \centering
  \includegraphics[width=\linewidth]{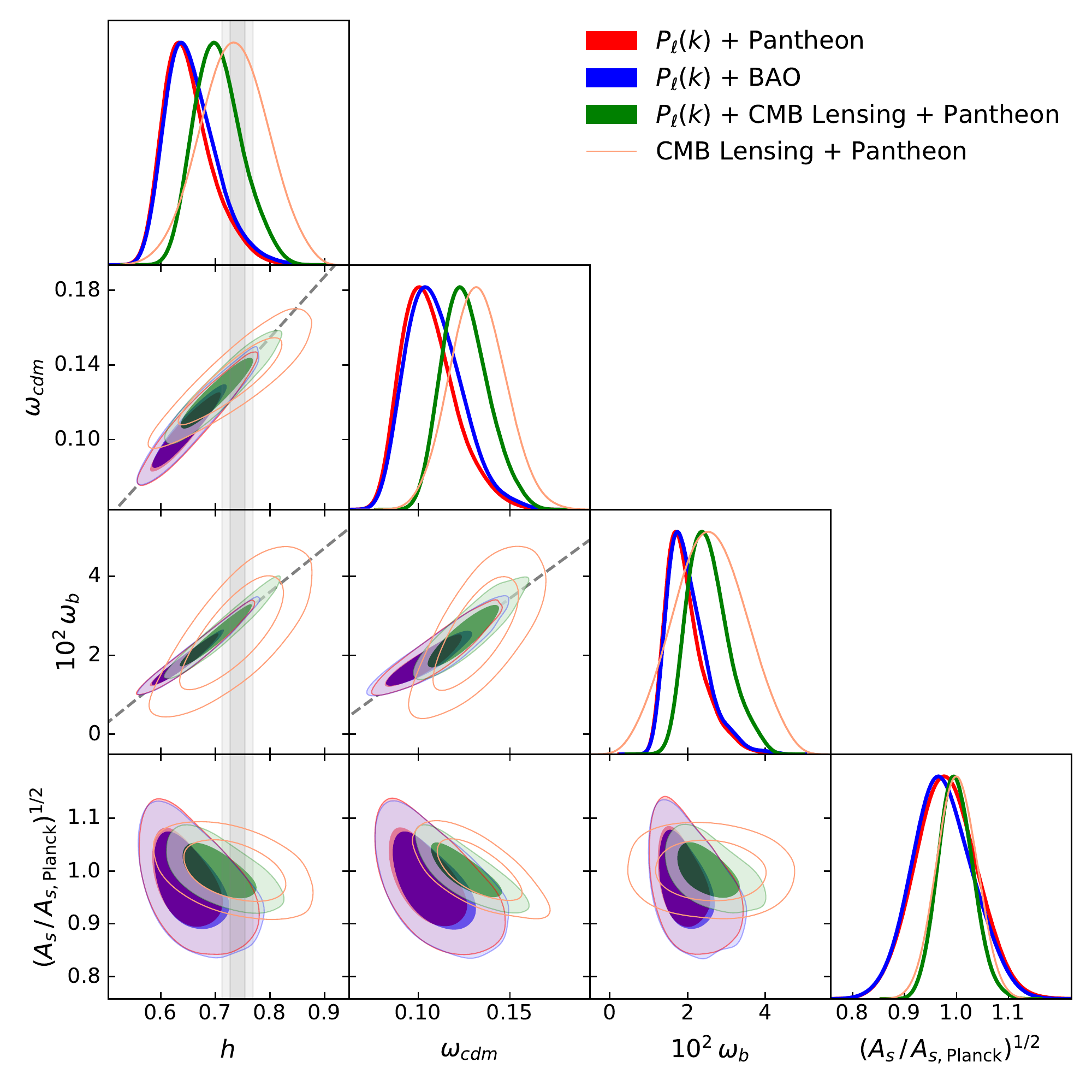}
  \caption{\small Parameter constraints from analyses of BOSS DR12 power spectra and \textit{Planck} lensing (as in Ref.\,\citep{2020arXiv200704007B}). %\bds{A precise external prior on the baryon density is not assumed for this analysis. Hence, the $H_0$ constraints do not derive information from the sound horizon, as evidenced by subsequent figures.} 
  Data-sets are combined with either Pantheon supernovae or uncalibrated BAO to provide $\Omega_m$ information, but no $\omega_b$ prior is assumed, such that the $H_0$ constraints do not derive information from the sound horizon (as evidenced by subsequent figures). Dashed lines show linear relationships and vertical bands give the SH0ES $H_0$ constraint \citep{2019ApJ...876...85R}. We omit the posteriors for $n_s$, $\sum m_\nu$ and the 12 nuisance parameters for clarity. \neww{Following the caption ordering}, $68\%$ $H_0$ confidence intervals are $65.1^{+3.0}_{-5.4}$, $65.6^{+3.4}_{-5.5}$, $70.6^{+3.7}_{-5.0}$ and $73.4\pm 6.1$ respectively, in $\hun$ units.
  %\blake{make clear which constraint correspond to which result / run.}}
  }
%   are also marginalized over, but omitted for clarity. }
%   Sound horizon independent parameter constraints from current data-sets, based on a full-shape analysis of the BOSS DR12 power spectra. We additionally include uncalibrated BAO data (from the BOSS DR7 6dF and main galaxy samples, the BOSS DR12 galaxies and the eBOSS DR14 Lyman-alpha catalogs) and measurements of CMB lensing (from the \textit{Planck} lensing likelihood). In all cases, we apply an $\Omega_m$ prior of $0.298\pm 0.022$ from the Pantheon supernova sample and a weak $\sim 8\%$ prior on $A_s$. To help with convergence, we use a broad $50\%$ prior on $\omega_b$. Dashed lines show linear relationships between parameters and we plot the \citet{2019ApJ...876...85R} $H_0$ measurement in vertical bands. Note that $n_s$ and $\sum m_\nu$ are also varied over, but are omitted for clarity. $1\sigma$ confidence intervals on $H_0$ are $65.1^{+3.0}_{-5.4}$, $65.9^{+3.2}_{-5.5}$ and $65.3^{+1.5}_{-2.9}$ for the three data-sets respectively.}
  \label{fig: obs-data}
\end{figure}%

\begin{figure}
  \centering
  \includegraphics[width=\linewidth]{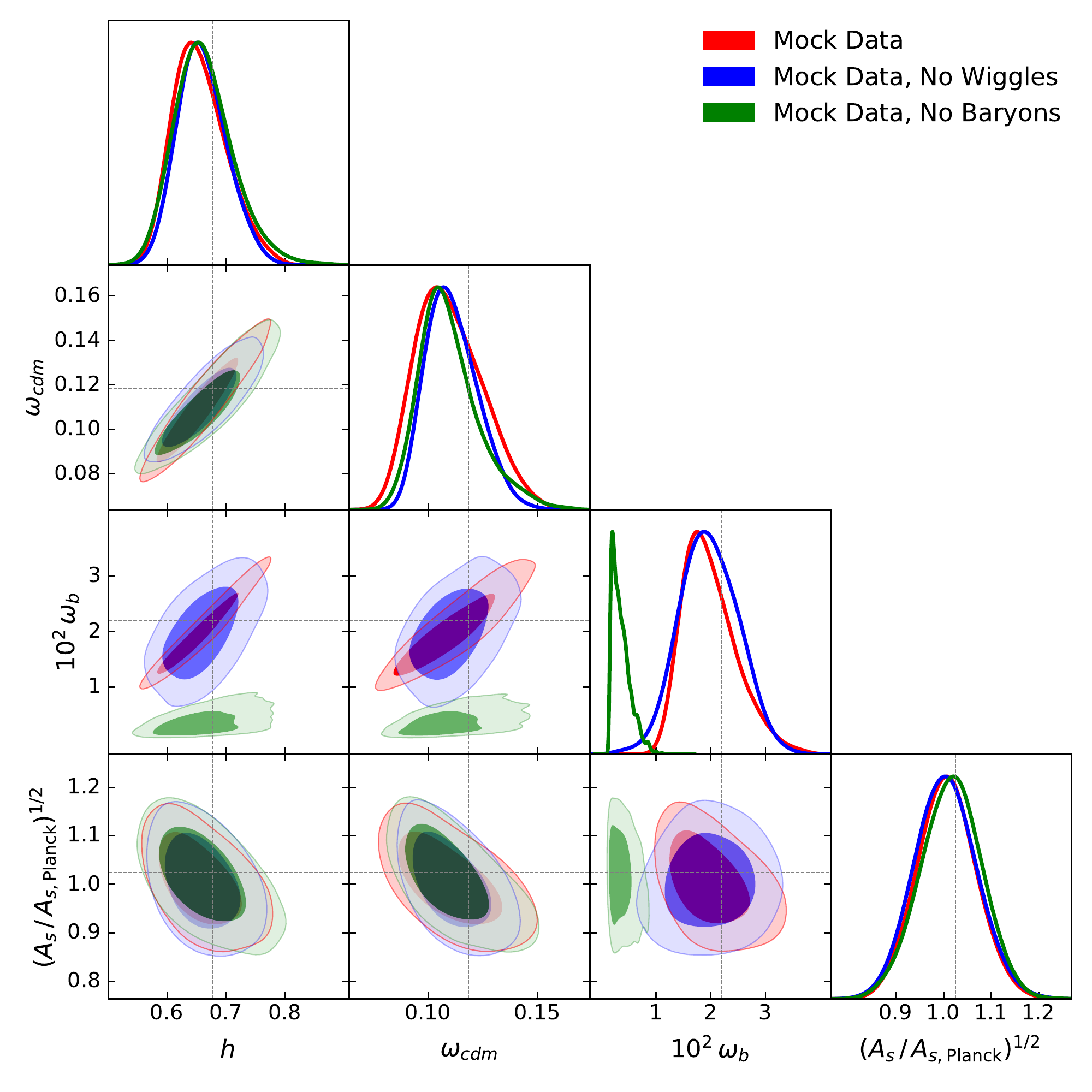}
  \caption{\small As Fig.\,\ref{fig: obs-data}, but for three mock data-sets: fiducial; with BAO wiggles removed; and with %in addition to the fiducial results, these include mock data with BAO wiggles removed and with 
  $\omega_b$ reduced by a factor of 10. All analyses include Pantheon priors on $\Omega_m$. 
  Thin lines show the true parameter values, and the $1\sigma$ constraints on $H_0$ are $65.6^{+3.7}_{-5.3}$, $65.9^{+3.7}_{-4.6}$ and $66.2^{+4.1}_{-5.6}$ in $\hun$ units. The insensitivity of $H_0$ constraints to the removal of BAO wiggles and baryon damping implies that information is not being sourced from the sound horizon scale.%\bds{The fact that the width of the $h$ posterior does not change when removing the BAO or baryon-derived information shows that constraints are not arising from the sound horizon scale.
  %}
  %\blake{for x, y and z respectively (make clear which constraint is which result)}.
  }
%  ; one matching the BOSS DR12 sample, one removing BAO wiggles from the spectra, and one reducing $\omega_b$ by a factor of 10. All data-sets are generated with the same $h$, $\omega_{cdm}$ and $A_s$. Thin lines show the true values of each parameter, and we obtain $1\sigma$ confidence intervals on $H_0$ of $65.6^{+3.7}_{-5.3}$, $66.0^{+3.7}_{-4.6}$ and $66.6^{+4.1}_{-5.6}$.}
  \label{fig: mock-data}
\end{figure}

\subsection{Constraints from current data-sets}\label{subsec: current-constraints}
Fig.\,\ref{fig: obs-data} shows the cosmological constraints obtained. %from current observational data. 
Combining BOSS power spectra with Pantheon $\Omega_m$ priors, 
%on $\Omega_m$, 
we obtain $H_0 = 65.1^{+3.0}_{-5.4}\,\hun$, below the best-fit SH0ES value at a 95\% confidence level (including non-Gaussianity of the posterior), even though our analysis \resub{is not based on the sound horizon}.%does not depend on sound-horizon physics.
\footnote{\newww{This corresponds to $k_\mathrm{eq} = \left(1.40^{+0.10}_{-0.14}\right)\times 10^{-2}\,\hMpc$.}} 
%\oliver{I do not agree with the below rephrasing, since it implies that we are only below SH0ES at 95% CL \textit{since} the posterior is non-Gaussian, which I don't think is true.}
%\bds{Even though our analysis does not depend on the sound-horizon, this lies below the best-fit SH0ES value \citep{2019ApJ...876...85R}; however, we note that the SH0ES best-fit lies at the $95\%$ confidence level of this measurement, due to significant non-Gaussianity of the posterior.}
%\blake{lets make sure not to overstate the discrepancy with riess, this is less than 2 sigma as we havent even included the riess error.}\{oliver: let's just quote the confidence level.}
%in tension \blake{would say something else} with the best-fit SH0ES value \citep{2019ApJ...876...85R} \new{at a $96\%$ confidence level (including non-Gaussianity of the posterior)}, even though our analysis does not depend on $r_d$. 
%To encapsulate non-Gaussian posteriors, this is best enumerated by a `probability-to-exceed' \blake{maybe should rephrase here- confidence level?}, here $7.6\%$ for the SH0ES best-fit value. 
As expected, the $\omega_b$ posterior is broad, since, unlike in previous analyses, we have not imposed a restrictive prior. There is a strong $h-\omega_{cdm}-\omega_b$ degeneracy, close to the expected linear relationship, rather than the $\omega_{cdm}\propto h^{4}$ scaling of $r_d$-calibration. The $\omega_b-\omega_{cdm}$ degeneracy indicates that a small amount of $\omega_{cdm}$ information comes from the BAO wiggles, though, as argued below, we do not expect this to inform our $H_0$ constraints. \newww{We note little dependence on the $A_s$ prior, with $< 10\%$ degradation in $\sigma_{H_0}$ if this is removed; this is expected since $A_s$ can be measured from the power spectrum through the loop corrections.}  %\blake{Does it? We just see the BAO degeneracy -- the marginalized constraints are super broad -- but the ruler is barely calibrated, right?}\oliver{The omega_b information does come from BAO, but that's not an H0 calibration. It's one of the few ways to get omega_b information (and changes in the no-wiggle analysis)}

% In Fig.\,\ref{fig: obs-data}, we show the constraints on key cosmological parameters for the observational data-sets discussed above. From the BOSS power spectrum alone, we obtain a somewhat competitive constraint on $H_0$; $65.4^{+3.2}_{-5.2}$, in tension with the results of \citet{2019ApJ...876...85R}, even though our analysis is centred around the equality scale and not the sound horizon. Since the posteriors are somewhat non-Gaussian we will quote this tension as a \textit{probability-to-exceed} (PTE), here $7.6\%$. As expected, the constraints on $\omega_b$ are narrow, since we have not imposed a restrictive prior, unlike previous works. It is this that makes our analysis independent of $r_d$. Evidence for this is shown in the degeneracy directions; we observe a close-to-linear relationship in the $\omega_{cdm}-h$ and $\omega_b-h$ planes, matching that expected from the simple relations discussed above,
% %in Sec.\,\ref{sec: physics}, 
% and not the strongly non-Gaussian bananas expected for the $r_d$-based $\omega_{cdm}\sim h^{4}$ scaling. Considering the $\omega_b-\omega_{cdm}$ plane, we find again a close-to-linear dependency; in this case this is an indication of $\omega_b$ being measured from the BAO wiggle amplitude (which is sensitive to $\omega_b/\omega_{cdm}$), but, as argued in the following subsection, we do not expect this to modify our $H_0$ contour.

Using uncalibrated BAO instead of the Pantheon sample gives a similar posterior; $H_0 = 65.6^{+3.4}_{-5.5}\,\hun$. This is unsurprising; the sound horizon rescaling parameter, $\alpha_{r_d}$, removes the $H_0$ information, and the marginalized $\Omega_m$ \neww{constraint} from BAO alone is similar to Pantheon.
%\eric{could be interesting to quote the $\Omega_m$ constraint from uncalibrated BAO here}\oliver{To add}. 
Interestingly, the $\alpha_{r_d}$ posterior is $0.993\pm 0.016$; the combination of equality-based power spectra and (independent) uncalibrated BAO prefer a sound horizon consistent with \lcdm.

%We consider two additional data-sets in Fig.\,\ref{fig: obs-data}; BAO and CMB lensing. Due to the insertion of a free sound horizon rescaling parameter $\alpha_{r_d}$, the BAO information cannot add $H_0$ information directly (as $H_0$ is fully degenerate with $r_d$), though information can be wrought about $\Omega_m$ from the distance-redshift relation, particularly from the quasars. In this case, we note no noticeable improvement from the addition of BAO data, indicating that the added information is subdominant to that contained within the full-shape $P(k)$ constraints (which probe $\omega_{cdm}$ and the Pantheon $\Omega_m$ prior. Higher precision data would be needed for this to have a noticeable impact, or a restrictive prior on $\alpha_{r_d}$. Indeed, the data itself constrains $\alpha_{r_d} = 0.9961\pm 0.0077$, implying that the combination of equality-based $P(k)$ measurements with uncalibrated BAO prefers a sound horizon consistent with that predicted by \lcdm.

Combination with \textit{Planck} lensing shifts the $H_0$ posterior to larger values, with a marginalized limit of $70.6^{+3.7}_{-5.0}\,\hun$. Due to the addition of galaxy information, this is somewhat tighter than the lensing-only constraints of Ref.\,\citep{2020arXiv200704007B}, though there is no improvement relative to the $P_\ell(k)$ posteriors, due to the broad error bars \neww{on the CMB lensing measurements} and similar degeneracy directions. Note that the lensing-only constraint \new{($H_0 = 73.4\pm 6.1\,\hun $)} is shifted somewhat from that of Ref.\,\citep{2020arXiv200704007B}, due to slightly different prior choices. %, in particular \newww{the wider prior on $A_s$}.%particularly for $A_s$ and the neutrino mass. % (which gives a slight downwards shift in the posterior without appreciably affecting its width). %\blake{add numerics and perhaps should discuss further.}

\subsection{Dependence on $r_d$}\label{subsec: mock-constraints}

%\blake{Add a few introductory sentences listing the 3? arguments we have for the lack of rd dependence, to guide readers through the coming paragraphs?}

We now demonstrate
%\neww{It is important %for our analysis to ensure 
that our $H_0$ constraints do not receive significant information from the sound horizon, using three tests: repeating the analysis on mock data-sets without baryon oscillations and damping; employing scale cuts; and performing Fisher forecasts, where we can explicitly marginalize over $r_d$.
%}

First, we turn to the synthetic data-sets discussed above. %, which also provide a useful consistency test, since they are generated from known cosmological parameters.
%To ensure that the above $H_0$ constraints are not directly affected by sound horizon physics, we first turn to the mock data-sets discussed above. Since these are generated from a known cosmology, they are also a useful consistency test. 
% Whilst the above section clearly demonstrates the constraints one can place on $H_0$ from current data-sets, one question remains; how can we be sure that we do not have dependence on the sound horizon $r_d$? Indeed, careful consideration of the above degeneracy directions, %Sec.\,\ref{subsec: current-constraints},
% will reveal that the degeneracy direction is not \textit{quite} a linear relationship, but slightly tilted towards a steeper slope. To investigate this, we take a multi-pronged approach first by considering the constraints obtained from the aforementioned mock data-sets.%described in Sec.\,\ref{subsec: pk-data}. 
% Since these are obtained from known cosmology, they additionally provide a useful consistency test.
As shown in Fig.\,\ref{fig: mock-data}, our $H_0$ constraints are negligibly impacted by removing BAO wiggles or reducing baryonic damping. Since the mock data are generated to match the BOSS spectra, this is a strong indication that our $H_0$ constraints are independent of sound horizon physics.\footnote{\resub{That the constraints are not affected by the removal of \textit{all} baryon information also indicates that the weak prior placed on $\omega_b$ in the real analysis is not affecting the $H_0$ posterior in the fiducial analysis.}} Note that the best-fit values of $\omega_{cdm}$ and $h$ are shifted \new{by $\sim 0.5\sigma$} from the truth;
%\blake{can we be more quantitative here? e.g. fiducial value, how far offset, how many sigma.}
%(though remain $1\sigma$ consistent); 
this indicates a (modest) prior-volume effect \resub{due to non-Gaussianity of the high-dimensional posterior}, confirmed by its removal when reanalyzing the data with a covariance appropriate for a $10\times$ larger survey. Whilst this could be ameliorated by stricter nuisance parameter priors, given that the offsets are small, we do not include these. % for the sake of generality. 
The mocks also highlight the importance of $\Omega_m$ priors; since the FS likelihood sources $\omega_{cdm}$ information from BAO wiggles, the no-wiggle constraints on $H_0$ would degrade if an external prior was not present.

% The posterior contours are shown in Fig.\,\ref{fig: mock-data}. Our first observation is that the peaks are somewhat shifted from the true values for $\omega_{cdm}$ and $h$, though always $1\sigma$ consistent. By reanalyzing the data-set with a covariance appropriate for a $10x$ larger survey, we have confirmed that these are simply prior volume effects, and could be reduced by tightening the priors on nuisance parameters, though we do not do this for generality. Note also that these spectra highlight the utility of including the Pantheon prior on $\Omega_m$; since the BAO-free models cannot constrain $\omega_{cdm}$ from amplitude of the wiggles, they will be less well constrained unless an external prior is provided. 

% The main conclusion from Fig.\,\ref{fig: mock-data} however is that the $H_0$ constraints are not significantly affected by reducing $\Omega_b$ or removing wiggles from the spectra, and indeed they are narrower when the spectrum does not contain the BAO feature. If information was coming from the BAO rather than equality, we would expect a strong degradation of $H_0$ constraints when the BAO wiggles and/or baryon suppression were removed.

\begin{figure}
  \centering
  \includegraphics[width=\linewidth]{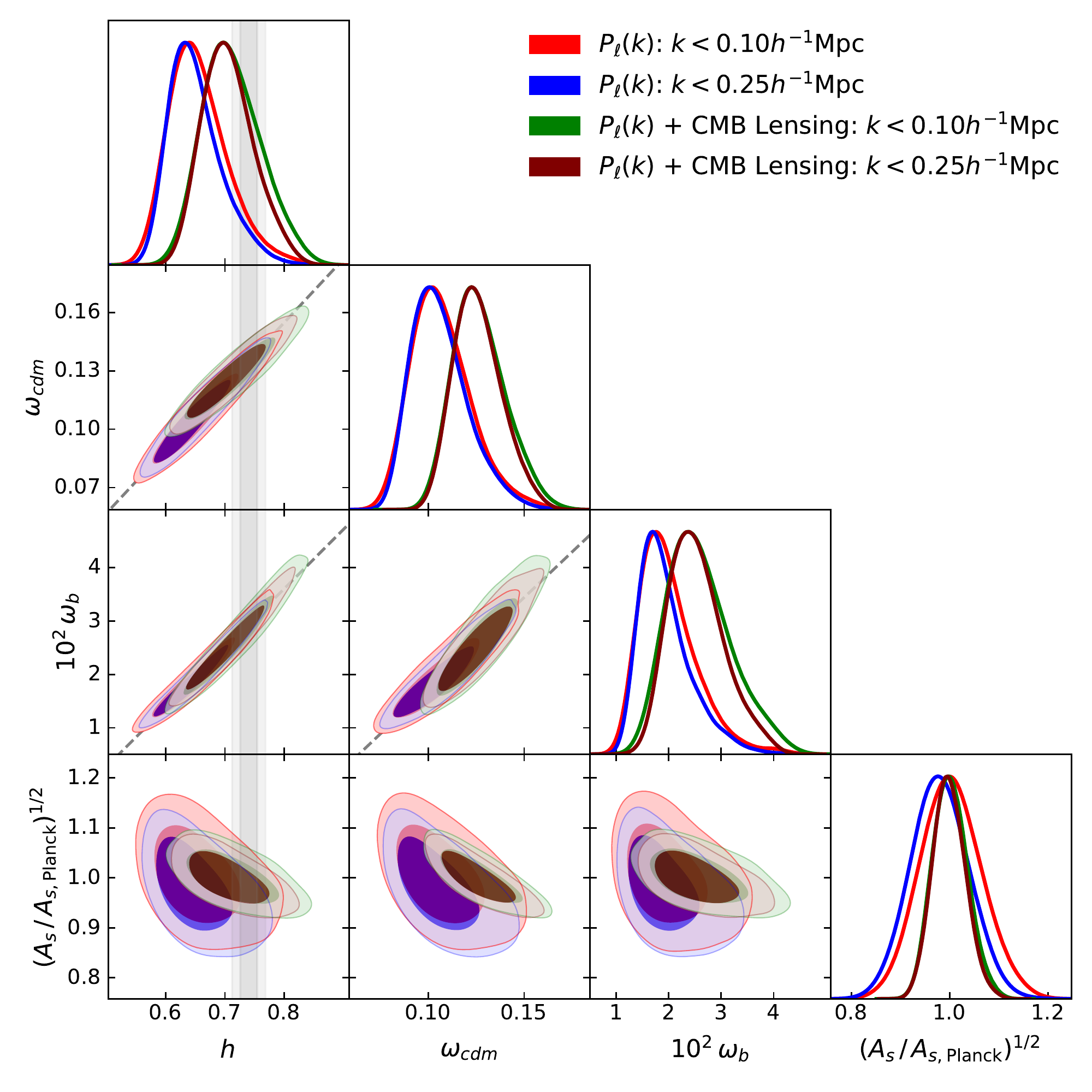}
  \caption{\small As Fig.\,\ref{fig: obs-data}, but restricting the $k$ range of the BOSS power spectrum analysis from the fiducial value of $k_\mathrm{max} = 0.25\hMpc$. All analyses include the Pantheon $\Omega_m$ prior. For the $k_\mathrm{max} = 0.1\hMpc$ data-sets, $H_0$ constraints are $65.6^{+3.8}_{-5.8}$ (BOSS) and $71.2^{+4.3}_{-5.7}$ (BOSS + \textit{Planck}) in $\hun$ units. The lack of dependence of $H_0$ on $k_\mathrm{max}$ again indicates that information is not arising from the BAO scale.%\bds{The insensitivity of $H_0$ constraints to scale cuts again indicates that information is not arising from the BAO scale.}
  }
  \label{fig: cut-data}
\end{figure}%

%Further evidence \new{that the sound horizon information is not contributing to the Hubble constraint}
%\eric{that the sound horizon information is not contributing to the Hubble constraint}
%s provided by scale-cuts. 
Scale cuts provide further evidence to support our conclusions. Fig.\,\ref{fig: cut-data} shows the effect of reducing $k_\mathrm{max}$ from $0.25\hMpc$ to $0.10\hMpc$, which, if information were coming from BAO wiggles, would be expected to \resub{significantly} inflate the $H_0$ posterior \resub{\citep[e.g.,][]{2020JCAP...05..042I}}. Notably, the reduction in constraining power is slight \new{($\sim 10\%$)}, though the nuisance parameters of the one-loop model suffer significant posterior inflation. Again, this indicates that the primary information is sourced by $k_\mathrm{eq}^{-1}$ rather than $r_d$ \resub{(and thus relatively large scales, though we note the lowest $k$-modes have limited impact due to their large statistical error)}.

%A second piece of evidence is obtained from performing scale-cuts. Whilst the main analysis in this work uses $k_\mathrm{max} = 0.25\hMpc$, in order to facilitate the inclusion of a well-constrained one-loop perturbation theory model, we may consider analyses with lower $k_\mathrm{max}$. If the bulk of the information comes from the equality scale, we expect the contours to be relatively insensitive to restricting to larger scales, though may inflate slightly since we lose full-shape information regarding $\omega_{cdm}$. Fig.\,\ref{fig: cut-data} shows this is indeed the case; there is no significant inflation of the error bars when $k_\mathrm{max}$ is reduced to $0.10\hMpc$, even though this barely contains significantly fewer BAO wiggles. A similar conclusion holds for the measurements including the CMB lensing constraints. There is notably some shift in $\omega_b$ and $A_s$ at low-$k$; this is expected since these receive information from the amplitude of BAO wiggles and one-loop terms respectively, and thus will be less well constrained using lower $k$ data.

\begin{figure}
  \centering
  \includegraphics[width=\linewidth]{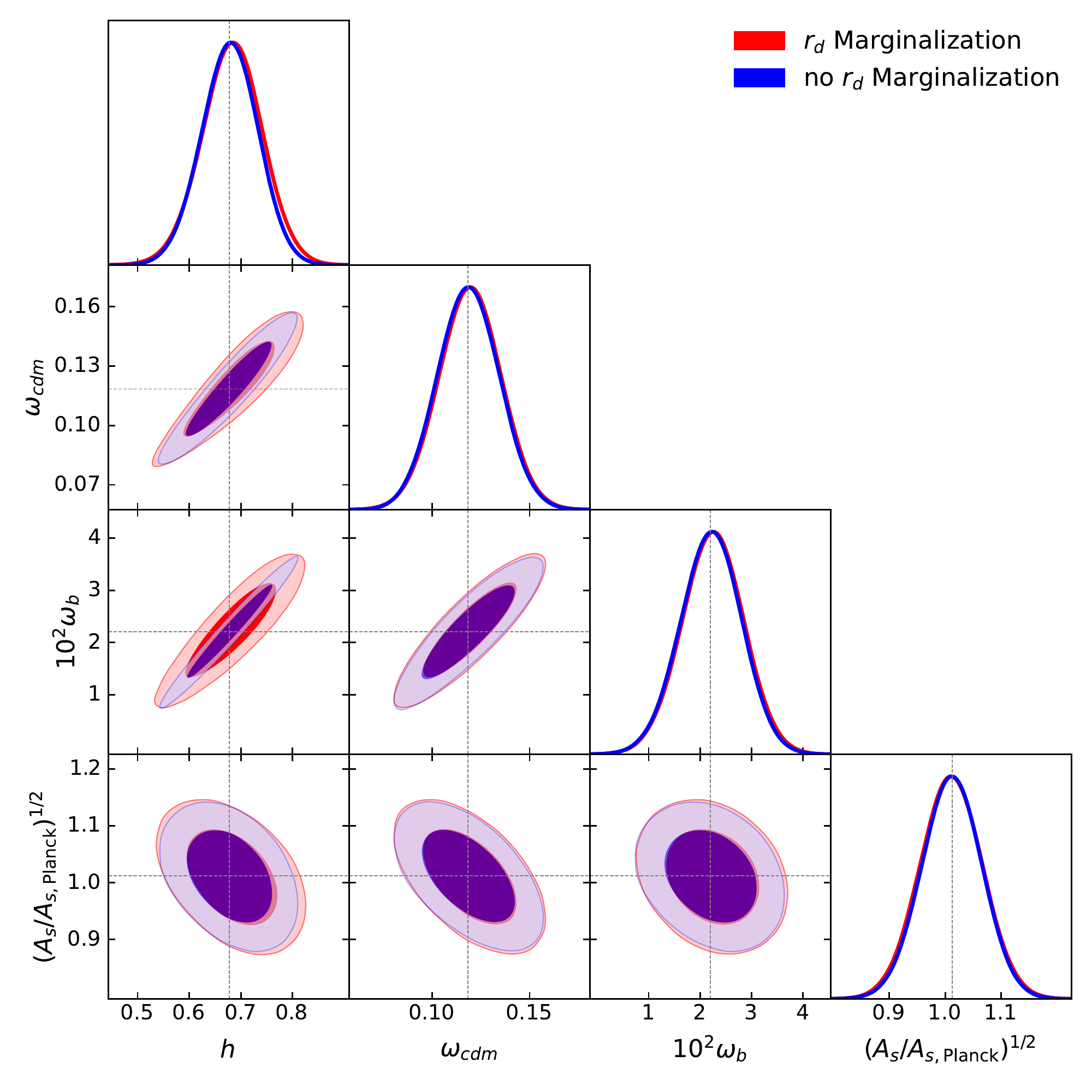}
  \caption{\small Posteriors from a simplified Fisher forecast mimicking the BOSS + Pantheon results, including explicit marginalization over the sound horizon, $r_d$, using the Eisenstein-Hu transfer function \citep{1998ApJ...496..605E}. 
  We obtain $\sigma_{H_0} = 5.9$ ($5.5$) $\hun $ with (without) $r_d$-marginalization; the small size of this degradation supplies further evidence that our constraints are insensitive to the sound horizon scale.}
  %\oliver{SHOULDN'T STATE MEANS FOR FISHER FORECAST}
  %$1\sigma$ $H_0$ constraints are $68.3\pm5.9$ ($67.8\pm5.5$) $\hun $ with (without) $r_d$-marginalization.
  %\oliver{MOVING THIS DISCUSSION INTO THE TEXT}
  %These have $1\sigma$ posteriors $H_0=68.3\pm5.9$ ($67.8\pm5.5$) $\hun $ with (without) marginalizing over $r_d$. \neww{The similarity of these constraints suggests that information about $r_d$ does not inform the $H_0$ constraint in our analysis. Calibration of $H_0$ via $k_{\rm eq}$ requires priors on $A_s$ and $\Omega_m$; removing these causes information about $H_0$ to enter via $r_d$ instead\gerrit{I think our understanding now is that this is actually more nuanced so maybe we should not say it like that. If somebody thinks very hard about it that might raise questions.}, thus \newww{without these priors} $r_d$-marginalization degrades the $H_0$ constraint significantly} ($H_0 = 68 \pm 11\,\hun$ compared to $H_0=67.8\pm5.7\,\hun$)\gerrit{If we have space we might want to include our argument for why this is the case in the main text}.}
  \label{fig:fisher_results}
\end{figure}%

Finally, we consider a simplified Fisher analysis in which $r_d$ can be marginalized over exactly. This is made possible by using the Eisenstein-Hu transfer function \citep{1998ApJ...496..605E} in \texttt{CLASS-PT}, rather than the usual output from \texttt{CLASS}. 
%For this, we have modified \texttt{CLASS-PT} to allow it to compute non-linear predictions for an arbitrary input linear power spectrum. 
In addition to the five cosmological parameters $\{h, \omega_{cdm}, \omega_b, A_s/A_{s,\mathrm{Planck}}, n_s\}$ (the Eisenstein-Hu approximation does not allow for massive neutrinos) we vary a sound horizon rescaling parameter $\beta_{r_d}$, \resub{which rescales $r_d$ within the transfer function}.
%\eric{called this $\alpha_{r_d}$ before?}
For simplicity, a single redshift \new{bin}
%\blake{bin}
(centered at $z = 0.51$) with the total BOSS volume is used, and window function effects are ignored. %otherwise
%ignore window function effects. 
As seen in Fig.\,\ref{fig:fisher_results}, the $H_0$ posteriors are broader than those found in Fig.\,\ref{fig: obs-data}; this is due to the assumptions of an Eisenstein-Hu model and exclusion of redshift evolution, yet the model retains qualitative utility. Marginalization over $r_d$ has little effect, reducing $\sigma_{H_0}$ by $< 10\%$, reinforcing our conclusions that the $H_0$ constraints are insensitive to $r_d$. Without an $\Omega_m$ prior the marginalization gives significant degradation, with $\sigma_{H_0} = 11\,\hun$; in this case, $\Omega_m$ information enters from the BAO wiggle \textit{amplitudes} which are washed-out by the marginalization.

\resub{The Fisher formalism may also be used to test the dependence of the $H_0$ constraints on the range of $k$-modes included in the analysis. Given that data systematics are concentrated in the first few $k$-bins, this is a useful probe of our sensitivity to such effects. Rerunning the above forecast (without marginalization over $\beta_{r_d}$) increasing $k_\mathrm{min}$ from $0.01\hMpc$ to $0.03\hMpc$ ($0.05\hMpc$) gives an $H_0$ posterior inflated by $14\%$ ($35\%$). We thus stress that the constraints found in this work have greater dependence on large-scale modes (and thus any present systematics) than for most BAO-only analyses. This is as expected since most of the $k_\mathrm{eq}$ information is wavenumbers in the linear regime, yet large enough to avoid excessive cosmic variance. We also note that the constraints are not affected by removal of the weak $\omega_b$ prior for $k_\mathrm{min} = 0.01\hMpc$ and $0.03\hMpc$, but suffer $\sim 10\%$ inflation if $k_\mathrm{min} = 0.05\hMpc$. This indicates that the results are prior-limited only if most of the large-scale power is removed.}

\subsection{Forecasting for Future Surveys}\label{subsec: future-forecast}
\newww{To estimate the potential of future surveys to constrain $H_0$ without the sound horizon, we perform a simplistic Fisher analysis, similar to that presented above. In particular, we consider a Euclid-like survey in eight redshift bins, taking the volumes and fiducial bias parameters from the forecast of Ref.\,\citep{Chudaykin2019}. For consistency, we slightly expand our $k$-range up to $k_\mathrm{max}=0.3\hMpc$ and do not impose nuisance parameter priors. Adopting the $A_s$ and $\Omega_m$ priors of this work, and marginalizing over $r_d$, we obtain $\sigma_{H_0}\sim 1.7\,\hun$; this tightens to $\sim 1.6$ $\hun$ with the more optimistic $\sigma_{\Omega_m} = 0.012$ prior of Ref.\,\citep{2020arXiv200704007B}. % ($\sigma_{\Omega_m}=0.012$).}
For future surveys, it is unclear whether removing the $\omega_b$ prior will be sufficient to ensure $r_d$-independence; this will be discussed in future work alongside a more complete forecast.}%more complex methods are required to marginalize over $r_d$%A full forecast is non-trivial however, and will require the development of a complete analysis method that is insensitive to $r_d$. This will be discussed in future work.

\section{Discussion}\label{sec: discussion}
In the past decade, galaxy surveys have focused on measuring BAO. In this work, we make use of the fact 
%\blake{would avoid claiming that this is the first measurement of keq - see olden-days shape parameter Gamma measurements by SDSS!} 
that an additional standard ruler is present; the horizon size at matter-radiation equality, $k^{-1}_\mathrm{eq}$. Combining galaxy power spectra from BOSS with cosmological priors on $\Omega_m$ gives equality-based constraints of $H_0 = 65.1^{+3.0}_{-5.4}\,\hun$ (power spectrum only) and $70.6^{+3.7}_{-5.0}\,\hun$ (adding \textit{Planck} lensing). For BOSS such a measurement can be obtained simply by analyzing the data without use of an informative $\omega_b$ prior; we demonstrate this using mock catalogs, scale cuts and Fisher forecasts. For the next generation of surveys, simple forecasts indicate that sound horizon independent constraints of \newww{$\sigma_{H_0} \simeq 1.6\,\hun$} should be possible; more sophisticated techniques may be required to remove $r_d$ information, however.

To close, we consider implications for the `Hubble tension'. Most proposed mechanisms for its resolution rely on modifying the sound horizon at recombination, and thus altering the BAO scale. Given that equality-based measurements are sensitive to higher redshifts than BAO measurements, $H_0$ constraints anchored at $z_d$ and $z_\mathrm{eq}$ may differ if new physics is at work, %\footnote{especially given that new physics models must also not break consistency of the sound horizon and equality scales}, 
making this a valuable test of new physics prior to recombination. Here, we find good agreement between galaxy-only $H_0$ measurements derived from the sound horizon \resub{\citep[e.g.,][]{2020JCAP...05..042I,2020JCAP...05..032P}} and equality scales, both of which favor lower values than those of SH0ES. If this consistency holds to much higher precision, it will \newww{place strong bounds on many} %make
beyond-\lcdm  %new-physics  
resolutions of the Hubble tension.\\ % difficult.

%If, with future measurements, a significant discrepancy arises between these two types of indirect $H_0$ measurements, it could indicate new physics prior to recombination; a consistency however, will make beyond-\lcdm resolutions of the Hubble tension difficult.
%. However, if the current consistency holds to much higher precision, it will make beyond-\lcdm  %new-physics  
%resolutions of the Hubble tension difficult.

%With the upcoming generation of galaxy surveys, the precision of such measurements will only increase; if the consistency holds, it will make explanation of the Hubble tensions with general beyond-\lcdm pre-recombination physics difficult.

%Most of the proposed mechanisms to alleviate this work by modifying the sound horizon at recombination, and hence altering the BAO scale. Since equality-based measurements are sensitive to higher redshifts, one would expect $H_0$ measurements based at $z_d$ and $z_\mathrm{eq}$ to differ if new physics were at work. In this work, we find good agreement between the $H_0$ measurements at the two redshifts; if this continues to higher precision, it may place another nail in the coffin of post-\lcdm physics.

\begin{acknowledgments}
\footnotesize
We thank Mikhail Ivanov, Julien Lesgourgues and Marko Simonovi\'c for insightful discussions; we are additionally grateful to Jo Dunkley, Dragan Huterer, Eiichiro Komatsu and Matias Zaldarriaga for comments on a draft of this manuscript, \resub{and to \resub{Adam Riess} and the anonymous referees for feedback on the original submission}. OHEP would like to thank the Max Planck Institute for Astrophysics for hospitality when this work was being finalized. OHEP acknowledges funding from the WFIRST program through NNG26PJ30C and NNN12AA01C.
GSF acknowledges support through the Isaac Newton Studentship and the Cambridge Trust Vice Chancellor's Award. BDS acknowledges support from an Isaac Newton Trust Early Career Grant, from a European Research Council (ERC) Starting Grant under the European Union’s Horizon 2020 research and innovation programme (Grant agreement No. 851274), and from an STFC Ernest Rutherford Fellowship.
\end{acknowledgments}

% \appendix
% \section{Dependence on $k_\mathrm{min}$}

% \begin{figure}
%   \centering
%   \includegraphics[width=\linewidth]{fisher_lowk.pdf}
%   \caption{\small \resub{Fisher forecast analyzing the dependency of the parameter posteriors on the truncation scale $k_\mathrm{min}$. As expected, the constraining power on $H_0$, $\omega_\mathrm{cdm}$ and $\omega_b$ is reduced when the largest modes are removed, with $\sigma_{H_0} = 4.7,\,5.0,\,5.9 \hun$ for the three contours respectively.}}
%   \label{fig:fisher_results}
% \end{figure}%

% \resub{We briefly comment on the dependence of the spectra on $k_\mathrm{min}$; the largest mode considered. For simplicity, we perform a Fisher analysis, as in Sec.\,\ref{subsec: mock-constraints}, analyzing the fiducial BOSS-like mock data-set, with $k_\mathrm{min}\in\{0.01,0.03,0.05\}\hMpc$. $k_\mathrm{max}$ is fixed to $0.25\hMpc$ in all cases, and we use the cosmological priors (including those from Pantheon) as above.}

\bibliographystyle{JHEP}
\bibliography{adslib,otherlib}% Produces the bibliography via BibTeX.

\end{document}